\documentclass[twocolumn,showpacs,superscriptaddress,email,floatfix,longbibliography]{revtex4-2}

\usepackage[left=2cm,right=2cm,top=2cm,bottom=2cm]{geometry}
\usepackage{amsmath}
\usepackage{inputenc}
\usepackage{graphicx}
\usepackage{braket}
\usepackage{longtable}
\usepackage{bbold}
\usepackage{amsmath,amssymb,amsfonts,bbm,graphicx,hyperref,times,color}
\usepackage{multirow}
\usepackage{graphicx}
\usepackage[table,xcdraw]{xcolor}
\usepackage{natbib}

\begin{document}
\title{Measurement-feedback control of chiral photon emission from an atom chain into a nanofiber}
\author{G. Buonaiuto}
\affiliation{Institut f\"ur Theoretische Physik, Universit\"at Tübingen, Auf der Morgenstelle 14, 72076 T\"ubingen, Germany}
\author{I. Lesanovsky}
\affiliation{Institut f\"ur Theoretische Physik, Universit\"at Tübingen, Auf der Morgenstelle 14, 72076 T\"ubingen, Germany}
\affiliation{School of Physics and Astronomy and Centre for the Mathematics and Theoretical Physics of Quantum Non-Equilibrium Systems, The University of Nottingham, Nottingham, NG7 2RD, United Kingdom}
\author{B. Olmos}
\affiliation{Institut f\"ur Theoretische Physik, Universit\"at Tübingen, Auf der Morgenstelle 14, 72076 T\"ubingen, Germany}
\affiliation{School of Physics and Astronomy and Centre for the Mathematics and Theoretical Physics of Quantum Non-Equilibrium Systems, The University of Nottingham, Nottingham, NG7 2RD, United Kingdom}

\begin{abstract}
We theoretically investigate measurement-based feedback control of a laser-driven one-dimensional atomic chain interfaced with a nanofiber. The interfacing leads to all-to-all interactions among the atomic emitters and induces chirality, i.e. the directional emission of photons into a preferred guided mode of the nanofiber. In the setting we consider, the measurement of guided light --- conducted either by photon counting or through homodyne detection of the photocurrent quadratures --- is fed back into the system through a modulation of the driving laser field. We investigate how this feedback scheme influences the photon counting rate and the quadratures of the guided light field. Moreover, we analyse how feedback alters the many-body steady state of the atom chain. Our results provide some insights on how to control and engineer dynamics in light-matter networks realizable with state-of-the-art experimental setups.
\end{abstract}

\maketitle

\section{Introduction}
Recent years have seen rapid progress in the development of experimental techniques and theoretical ideas concerning the real-time manipulation of quantum optical many-body systems \cite{mekhov2012,ritsch2013,hammerer2004,chou2005,browaeys2020,reiserer2015,duan2010,soare2014,hammerer2010}. This was partly motivated by potential applications in the realms of quantum computation and simulation \cite{barrett2005,sangouard2011,georgescu2014}. Feedback protocols have been identified as a promising strategy to control the dynamics, the stationary state, and the properties of light emitted from quantum optical systems \cite{sayrin2011,nelson2000,balouchi2017,vollbrecht2009,blok2014,grimsmo2014}. One of these protocols is measurement-based feedback, which entails the continuous application of control fields whose strength depends on the outcome of measurements performed on an emitted light field \cite{wiseman1994a,wiseman2009,jacobs2014,lammers2016,nurdin2017,zhang2017}. In essence, measurement-feedback control provides a set of prescriptions for the manipulation of open quantum systems, which can be exploited for driving the system into a desired state or creating a light source with specific properties. Different kinds of measurements, such as photon counting or homodyne detection, have been proposed theoretically as a method for cooling levitated heavy particles \cite{walker2019}, stabilizing the spatial mode for a trapped BEC \cite{szigeti2009}, and creating squeezed light states \cite{wiseman1993,wiseman1994} as well as strongly correlated states of atoms in a cavity \cite{mazzucchi2016}, among others. Successful experimental implementations of such methods include the improvement of trapped ion cooling \cite{bushev2006}, the suppression of quantum noise \cite{inoue2013} and thermal decoherence \cite{wilson2015}, as well as the control of correlations in a nanomechanical oscillator \cite{sudhir2017}, and the generation of entanglement between qubits \cite{riste2013}.

\begin{figure}[t]
\centering
\includegraphics[width=\columnwidth]{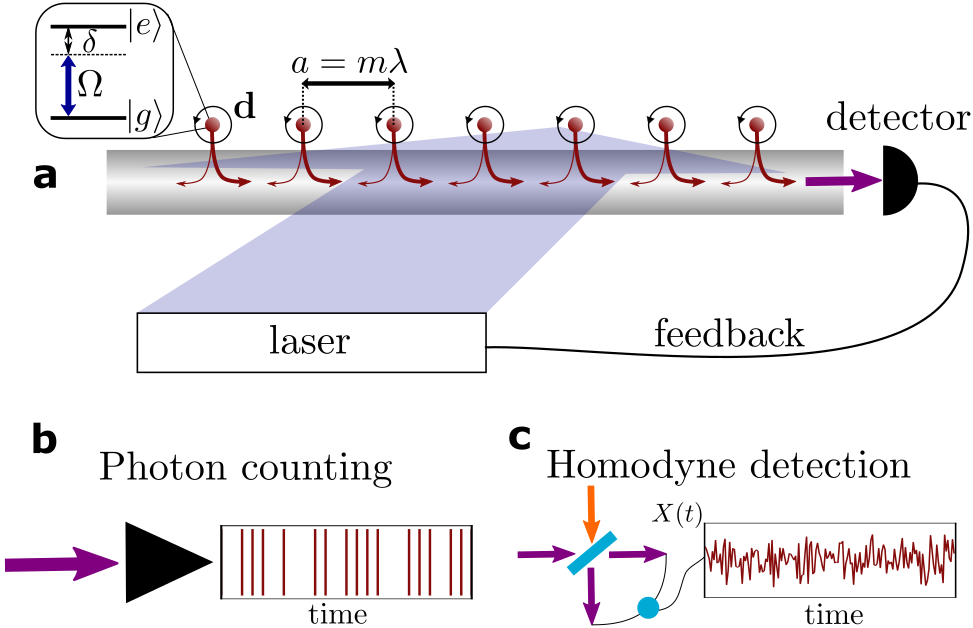}
\caption{\textbf{Atom-waveguide setup.} \textbf{a:} A chain of two-level atoms is placed in the vicinity of a waveguide (here a nanofiber). The atoms are driven homogeneously via a laser field with Rabi frequency $\Omega$ and detuning $\delta$ and they are separated from each other by a distance $a$ commensurate with the wavelength $\lambda$ of the laser field. The photons are emitted into the guided modes of the nanofiber. The measurement realized on the right-propagating photons feeds back into the amplitude the external laser field. The feedback is realized considering two possible detection schemes: \textbf{b:} Photon counting, where photons are detected at discrete times, triggering a finite laser pulse, or \textbf{c:} homodyne detection, where a quadrature of the photocurrent [here $X(t)$] is measured continuously. The external laser field is then modulated according to the measurement outcome.}\label{fig1}
\end{figure}

In this work, we explore how to use measurement-feedback to control the stationary state and the light emission properties from an atomic chain coupled to a one-dimensional nanophotonic waveguide. Nanophotonic waveguides \cite{mitsch2014,coles2016,bliokh2015} (here we will consider in particular a nanofiber) support a small number of guided electromagnetic modes, through which light propagates only longitudinally. By selecting appropriately the atomic positions and polarizations, one can open a photon decay channel into this set of modes. The guided fiber-modes induce all-to-all exchange interactions between the atoms and furthermore the atom-light coupling may be chiral, i.e. photon emission takes place preferably into one of the modes travelling into a given direction. These features make the atom-nanofiber setting a promising candidate for the realization of photonic quantum technologies \cite{lodahl2015,ruostekoski2016,guimond2016,asenjo2017,mahmoodian2017,lang2017,lodahl2017,buonaiuto2019,zhang2020,jen2020,jones2020,olmos2020}. Here, we consider a scenario where a continuous measurement of the light emitted into one of the guided modes triggers the feedback. We focus on two measurement schemes: the detection of single photons arriving at a photon counting detector, and the measurement of the optical light quadratures via homodyne detection \cite{walls2007,davidovich1996}. We investigate how this allows to manipulate the stationary many-body state of the atoms as well as the intensity and fluctuations of the chirally emitted light. The paper is structured as follows: in Section \ref{sec:system} we introduce the many-body master equation model describing the coupled atom-waveguide system. The measurement-based feedback protocol and the impact of the choice of the particular feedback protocol on the form of the many-body master equation is discussed in Section \ref{sec:feedback}. In Section \ref{sec:sensemble} we summarize the so-called $s$-ensemble method, which we use for calculating the full statistics of the photon emission. Finally, in Section \ref{sec:results} we present the main results of this work, including the analysis of the many-body stationary state, the photon counting statistics and the generalized optical quadratures. Conclusions and an outlook are found in Section \ref{sec:conclusions}.

\section{Atom-waveguide system without feedback} \label{sec:system}

The setup we consider here is sketched in Fig. \ref{fig1}a. It consists of a chain of atoms (modelled as two-level systems) with nearest neighbor distance $a$, which is parallel and in close vicinity to the longitudinal axis of a nanofiber. An external laser field with wave vector perpendicular to the chain drives the atoms with Rabi frequency $\Omega$ and detuning $\delta$. Photons are emitted from the chain into two \textit{guided} modes of the nanofiber, which are left- and right-propagating, respectively. There is, moreover, the possibility of photon emission into unguided modes of the electromagnetic field. 

For modelling the dynamics of this open quantum system, we follow the approach set out, e.g. in \cite{stannigel2012,ramos2014,pichler2015,guimond2016,buonaiuto2019}, which makes use of the Born-Markov approximation and assumes that the spacing between the atoms, $a$, is commensurate with the wavelength of the laser field $\lambda$, i.e. $a=m\lambda$ with $m=1,2,\dots$ (see Fig. \ref{fig1}a). This leads to a master equation (ME), which has a particularly simple form (we set $\hbar=1$):
\begin{equation}
\label{ME}
\dot{\rho}=-\mathrm{i}\left[H,\rho\right] +\gamma\mathcal{D}(J)\rho+\Gamma\sum_{j=1}^N\mathcal{D}(\sigma_j)\rho.
\end{equation}
The first term of this equation is determined by the Hamiltonian
\begin{eqnarray}
\label{Ham}
H&=&\Omega\left(J+J^{\dagger}\right)+\delta \sum_{j=1}^N\sigma_{j}^{\dag}\sigma_{j}\\
&&-\frac{\mathrm{i}}{2}\Delta\gamma \sum_{j>l}(\sigma^{\dag}_{j}\sigma_{l}-\sigma^{\dag}_{l}\sigma_{j}).\nonumber
\end{eqnarray}
Here, the first row describes the excitation of the atoms by a spatially uniform laser field, with $J=\sum_{j=1}^N\sigma_{j}$, and $\sigma_{j}=\left|g_j\right>\!\left<e_j\right|$ being the ladder operator connecting the two levels $\left|g_j\right>$ and $\left|e_j\right>$ of the $j$th atom. The second row describes exchange interactions among the atoms, which occur at a rate given by the parameter $\Delta\gamma=\gamma_{R}-\gamma_{L}$, where $\gamma_R$ and $\gamma_L$ represent the single-atom decay rate into the right- and left-propagating guided modes, respectively. Hence, an asymmetry of the photon emission rates --- so-called \emph{chirality} --- also leads to all-to-all interactions between the atoms. This can be controlled by an appropriate choice of the laser polarization and transition dipole moment $\mathbf{d}$ of each atom. In particular, the dipole moments are required to have a real and imaginary part (i.e. elliptically polarized light) for the coupling to the guided modes to acquire a chiral character, i.e. to have $\Delta\gamma\neq 0$ \cite{bliokh2015,mitsch2014}.

The second term of the ME (\ref{ME}) describes the incoherent emission of photons into both guided modes with total decay rate $\gamma=\gamma_{R}+\gamma_{L}$, making use of the superoperator
\begin{eqnarray*}
\mathcal{D}(J)\bullet=J\bullet J^\dag-\frac{1}{2}\left\{J^\dag J,\bullet\right\}.
\end{eqnarray*}
The last term of the ME \eqref{ME} captures the emission of photons into the unguided modes at rate $\Gamma$. Note, that while the emission into the guided modes is collective, i.e. the jump operator that describes such emission event is $J$, a symmetric superposition of all single-atom ladder operators, the emission into the unguided modes is considered here to be independent for each atom. Moreover, the dipole-dipole interactions induced by the free-space electromagnetic field are altogether neglected. These two last approximations are well justified due to the large values of the inter atomic separation $a$ that we consider here. Furthermore, since the condition $a=m\lambda$ is satisfied (see Fig. \ref{fig1}a), the emission into the unguided modes becomes less and less important compared to the emission of photons into the nanofiber as we increase the atom number $N$ \cite{lekien2008,jones2020}. For the sake of simplicity, we will from now on consider the case $\Gamma=0$ (photon emission takes place exclusively into guided modes), such that the ME reads
\begin{equation}\label{ME2}
    \Dot{\rho}=-\mathrm{i}\left[H,\rho\right]+\gamma_R\mathcal{D}(J)\rho+\gamma_L\mathcal{D}(J)\rho.
\end{equation}
For convenience, we separate here explicitly the emission of right- and left-propagating photons.

\section{Feedback} \label{sec:feedback}

The feedback protocol that we consider is shown in Fig. \ref{fig1}a: the light emitted into the right-propagating guided mode is analyzed, either by counting the photons that arrive at the detector at discrete times (Fig. \ref{fig1}b), or by continuous homodyne detection of a given quadrature of the photocurrent (Fig. \ref{fig1}c). The results of these measurements are fed back into the atomic system. In particular, the amplitude of the driving laser field is modified according to the stochastic measurement outcomes. This gives rise to a modified ME and dynamics, which leads to a stationary state with properties that can differ dramatically from the ones under the unconditional evolution that is governed by Eq. (\ref{ME2}). The feedback mechanism is assumed to be instantaneous, meaning that the delay time between the measurement and the application of the control field is small compared to the typical timescales of the atomic system. This assumption ensures the Markovianity of the dynamical description, thus allowing to write down a Lindblad master equation inclusive of the feedback effect. In the following, we summarize briefly how the feedback strategies alter the structure of the ME. A general overview can be found in the literature, e.g. in Refs. \cite{wiseman1993,wiseman2009,lammers2016}.

\subsection{Photon counting}

We first consider the case of photon counting (see Fig. \ref{fig1}b). Here, the detector clicks every time it registers a single photon that is emitted into the right-propagating mode of the nanofiber. Given that the feedback is instantaneous after the detection of a photon, the density matrix is evolved with the operator $e^{\mathcal{K}}$ immediately after the associated quantum jump, with $\mathcal{K}$ being a superoperator. This prescription results in a Markovian ME, which is written as \cite{wiseman2009}
\begin{equation}
\dot{\rho}=-\mathrm{i}\left[H,\rho\right]+\gamma_{R}e^{\mathcal{K}}J\rho J^{\dagger}+\gamma_{L}J\rho J^{\dagger}-\frac{\gamma}{2}\{J^{\dagger}J,\rho \}.
\end{equation}
We consider that the detection event triggers a laser pulse with a fixed area $g$, which in practice can be implemented by a fast modulation of the laser that excites the atoms. The corresponding feedback operation is unitary, reading $\mathcal{K}\bullet=-\mathrm{i}g\left[J+J^\dag,\bullet\right]$. Considering this, the ME simplifies to
\begin{equation}
\label{countingfb}
\dot{\rho}=-\mathrm{i}\left[H,\rho\right]+\gamma_{R}J_R\rho J_R^{\dagger}+\gamma_{L}J\rho J^{\dagger}-\frac{\gamma}{2}\{J^{\dagger}J,\rho \},
\end{equation}
where the jump operator corresponding to a photon emission into the right-propagating mode acquires the form $J_R=e^{-\mathrm{i}g(J+J^\dag)}J$.

\subsection{Homodyne detection}

The second feedback strategy is based on the homodyne photocurrent (see Fig. \ref{fig1}c): photons emitted into the right-propagating mode are analyzed via a homodyne detection scheme, which allows to measure the generalized optical quadratures of the output field:
\begin{equation}
X_\alpha(t)=\sqrt{\gamma_R}\left<e^{-\mathrm{i}\alpha}J+e^{\mathrm{i}\alpha}J^\dag\right>_t+\xi(t).
\end{equation}
Here, $\left<\cdot\right>_t$ represents the expectation value as a function of time $t$, and $\xi(t)$ represents real white noise with $\left<\xi(t^\prime)\xi(t)\right>=\delta(t-t^\prime)$. From this general expression one obtains for $\alpha=0$ and $\alpha=\pi/2$ the amplitude, $X(t)$, and phase quadrature, $P(t)$, of the light. Based on the measured quadrature we perform a modulation of the laser field exciting the atoms. This is represented by the superoperator 
\begin{equation}
    {\cal F}\bullet=-\mathrm{i}\sqrt{\gamma_R}g X_\alpha(t)\mathrm{d}t\left[F,\bullet\right],
\end{equation}
which acts at each infinitesimal time $\mathrm{d}t$. Here, $F=J+J^\dag$ and the dimensionless parameter $g$ represents the strength of the feedback. By keeping terms only up to first order in $\mathrm{d}t$ and taking the ensemble average \cite{wiseman1993,wiseman2009,lammers2016}, this results in a Markovian ME of the form
\begin{eqnarray}
\label{homodynefb}
    \dot{\rho}&=&-\mathrm{i}\left[H+\frac{g\gamma_R}{2}\left(e^{-\mathrm{i}\alpha}F^\dagger J+e^{\mathrm{i}\alpha}J^\dagger F \right),\rho\right]\\
    &&+\gamma_R\mathcal{D}(J-\mathrm{i}ge^{\mathrm{i}\alpha}F)\rho+\gamma_L\mathcal{D}(J)\rho.\nonumber
\end{eqnarray}
Note that, unlike in the case of photon counting, the feedback does not only affect the jump operator corresponding to the emission of right-propagating photons, but also changes the unitary dynamics.

\section{S-ensemble formalism} \label{sec:sensemble}

To characterize the properties of the light emitted from the nanofiber we use a formalism that is based on the $s$-ensemble approach to open quantum systems, details of which can be found, e.g. in Refs. \cite{garrahan2010,garrahan2011,hickey2012,ates2012,buonaiuto2019,olmos2020}. This method allows to access not only the average values but also the fluctuations and higher moments of time-integrated observables, such as the photon emission count and the time-integrated quadrature of the light emitted into the right-propagating mode. 

We first discuss photon counting: the probability of detecting the arrival of $K$ photons after an observation time $t$, assuming perfect detection efficiency, is given by $P_t(K) = {\rm Tr} \left[ \rho^{(K)}(t) \right]$. Here, $\rho^{(K)}(t)$ is the density matrix conditioned on having detected exactly $K$ photons. Large deviation theory \cite{eckmann1985,touchette2009} establishes that at very long times, the probability $P_t(K)$ acquires the so-called large deviation form $P_t(K) \approx e^{-t \varphi(K/t)}$, where $\varphi(K/t)$ is the large deviation function. The associated moment generating function $Z_t(s)$ can be shown to also have a large deviation form: $Z_t(s) \approx e^{t \theta_K(s)}$, where $s$ is a counting field conjugate to $K$, and $\theta_K(s)$ is the scaled cumulant generating function. In practice, the latter may be found by calculating the eigenvalue with the largest real part of a deformed master operator, ${\cal W}_s$. The form of this superoperator depends on the actual measurement. In the case of photon counting, it reads
\begin{eqnarray}
{\cal W}_s(\bullet) &=& -\mathrm{i} [ H , \bullet] + e^{-s}\gamma_{R}J_R \bullet J_R^{\dagger}\\
&&+\gamma_{L}J\bullet J^{\dagger}-\frac{\gamma}{2}\{J^{\dagger}J,\bullet \}.\nonumber
\label{Ws}
\end{eqnarray}
Once $\theta_K(s)$ is obtained here, its derivatives with respect to $s$ evaluated at $s=0$ yield the moments of the photon count distribution, such as the photon count rate
\begin{equation}
\label{counting}
k \equiv \frac{\langle K \rangle_{s=0}}{t} =  -\partial_s\theta_K(s)\big\rvert_{s=0},
\end{equation}
and the corresponding variance
\begin{equation}
    \Delta k^2\equiv\frac{\langle K^2 \rangle_{s=0}-\langle K \rangle_{s=0}^2}{t}=\partial^2_s\theta_K(s)\big\rvert_{s=0}.
\end{equation}

When the output detected is a quadrature of the photocurrent via homodyne measurements, $X_\alpha(t)$, the corresponding superoperator is \cite{hickey2012,olmos2020}
\begin{eqnarray}\label{sup2}\nonumber
{\cal W}_s(\bullet)&=&-\mathrm{i}\left[H+\frac{g\gamma_R}{2}\left(e^{-\mathrm{i}\alpha}F^\dagger J+e^{\mathrm{i}\alpha}J^\dagger F \right),\bullet\right]\\
    &&+\gamma_R\mathcal{D}(J-\mathrm{i}ge^{\mathrm{i}\alpha}F)\bullet+\gamma_L\mathcal{D}(J)\bullet\\\nonumber
    &&-\frac{s}{2}\sqrt{\gamma_R}\left[e^{-\mathrm{i}\alpha}(J-\mathrm{i}ge^{\mathrm{i}\alpha}F){\bullet}+\mathrm{h.c.}\right]  +\frac{{s}^{2}}{8}{\bullet},
\end{eqnarray}
with $F=J+J^\dag$. In this case, the scaled cumulant generating function $\theta_{X_\alpha}(s)$ provides the cumulants associated to the time-integrated photocurrent quadrature, $X_\alpha=\int_0^t \mathrm{d}X_\alpha(\tau)$, e.g. the average photocurrent quadrature per unit time $x_\alpha\equiv \frac{\langle X_\alpha \rangle_{s=0}}{t}$ and its variance $\Delta x_\alpha^2\equiv\frac{\langle X_\alpha^2 \rangle_{s=0}-\langle X_\alpha \rangle_{s=0}^2}{t}$.

\section{Results} \label{sec:results}

\begin{figure}
\includegraphics[width=\columnwidth]{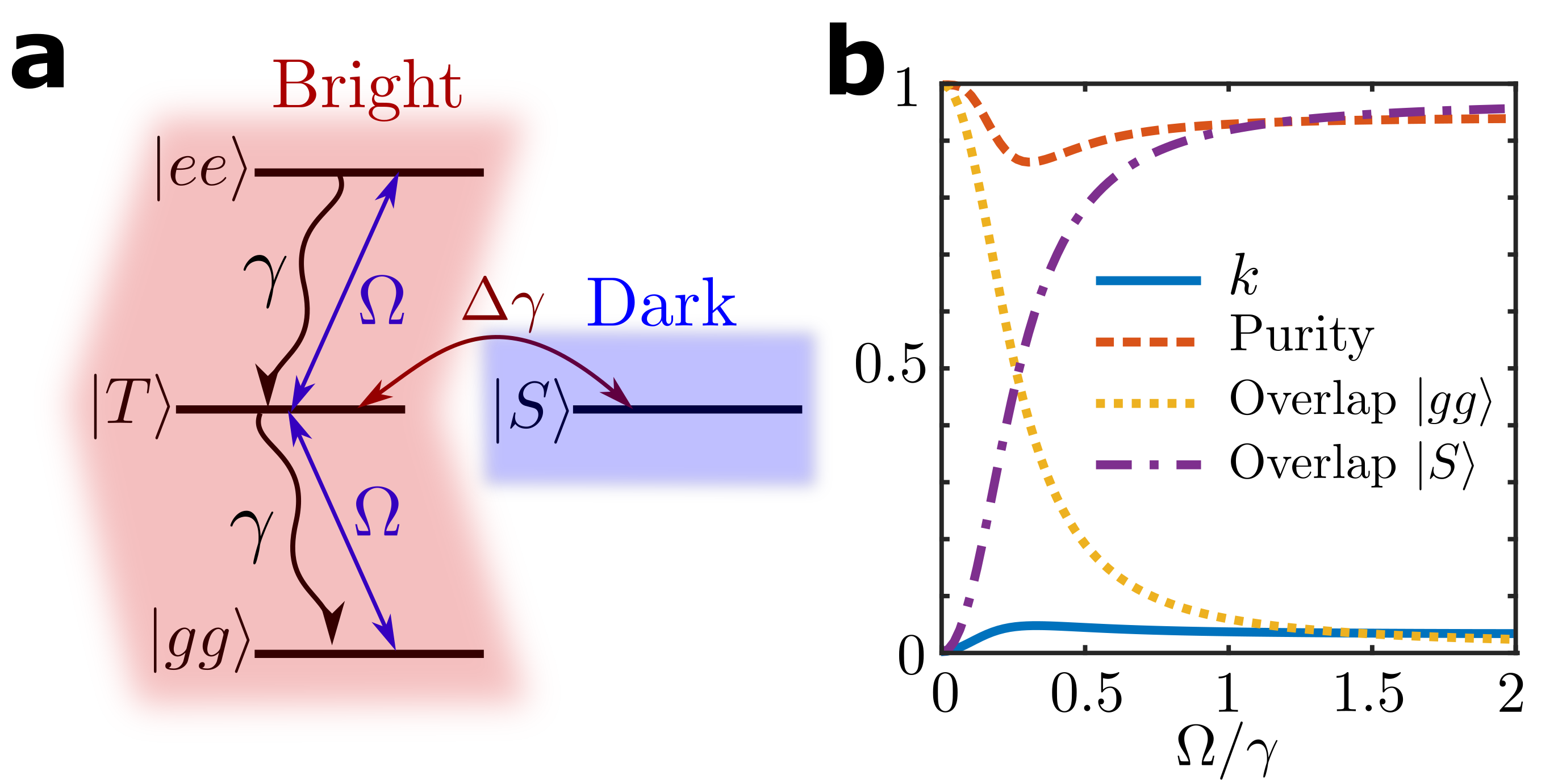}
\caption{\textbf{Two atoms without feedback.} \textbf{a:} Level scheme for resonant laser excitation, $\delta=0$. \textbf{b:} Photon counting rate $k$ (in units of $\gamma$), purity and overlap with $\left|gg\right>$ and $\left|S\right>$ of the stationary state for $N=2$ atoms. There is no feedback ($g=0$), the detuning of the laser excitation is $\delta=\gamma/10$, and the chirality is $\Delta\gamma=0.6\gamma$.}\label{fig2}
\end{figure}

We begin by considering a chain of $N=2$ atoms interfaced with the nanofiber in the absence of feedback. As shown in Fig. \ref{fig2}a, the system can be described in terms of four collective states: $\left|gg\right>$, $\left|T\right>=\frac{1}{\sqrt{2}}\left(\left|ge\right>+\left|eg\right>\right)$ (triplet state), $\left|S\right>=\frac{1}{\sqrt{2}}\left(\left|ge\right>-\left|eg\right>\right)$ (singlet state), and $\left|ee\right>$. The singlet state cannot be directly excited by the laser. It is only reached dynamically via the dipole-dipole exchange term in Eq. (\ref{Ham}), whose strength is controlled by the chirality $\Delta\gamma$. This state is \emph{dark}, i.e. it is not coupled via dissipation to any other state. Conversely, the remaining states, $\left|gg\right>$, $\left|T\right>$, and $\left|ee\right>$, are connected to each other via the laser field and the dissipation, and hence photons are emitted when the system is in this \emph{bright} subspace.

When the laser is on resonance, i.e. $\delta=0$, the stationary state of this system is a pure so-called dimer state \cite{stannigel2012,ramos2014,pichler2015,guimond2016,buonaiuto2019}
\begin{equation}
\left|D\right> = \frac{1}{\sqrt{\Delta \gamma^2+8\Omega^2}}\left[\Delta\gamma\left|gg\right>+\mathrm{i}2\sqrt{2}\Omega\left|S\right>\right].
\end{equation}
This state is also dark, i.e. no photons are emitted, independently of the value of the driving strength $\Omega$. Conversely, for $\delta\neq 0$, i.e. an off-resonant laser, the stationary state and corresponding emission rate depend on the value of $\Omega$. This is shown in Fig. \ref{fig2}b and discussed in detail in Ref. \cite{buonaiuto2019}. When the driving is weak, i.e. $\Omega\ll\gamma$, the (pure) stationary state of the system is simply the ground state $\left|gg\right>$. As $\Omega$ is increased, the stationary state becomes a "mixture" between the dark dimer state and a mixed one formed by the states of the bright subspace, such that the photon count rate $k$ becomes finite. This picture can be extended to larger system sizes, provided that the number of atoms, $N$, is even. Here, a many-body dark state emerges, which is a product of dimers, i.e. $\otimes_{j=1}^{N/2}\left|D\right>_{2j-1,2j}$.

\subsection{Photon counting feedback}

\begin{figure}
\includegraphics[width=\columnwidth]{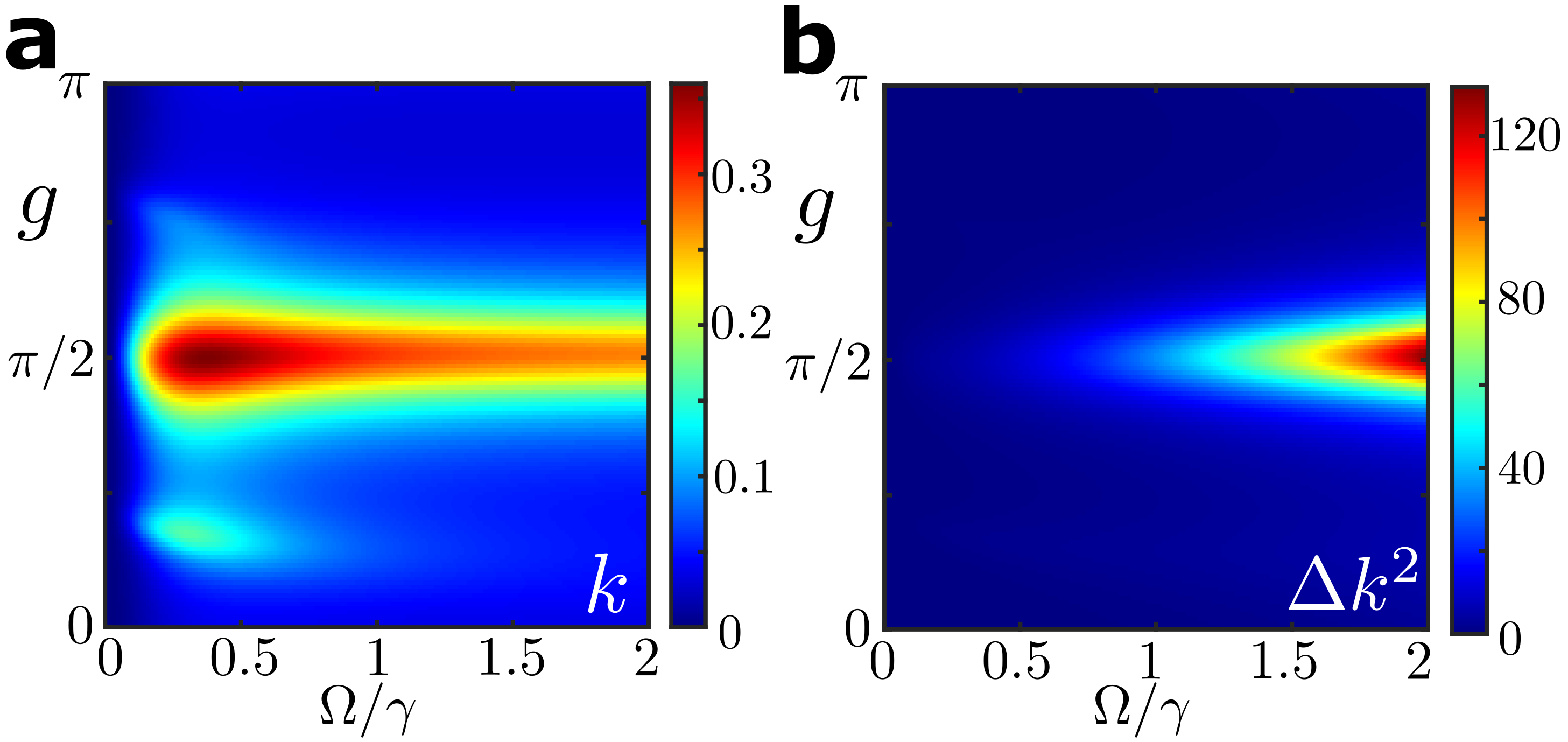}
\caption{\textbf{Photon counting feedback.} \textbf{a:} Photon counting rate $k$ and \textbf{b:} its fluctuations ${\Delta k}^2$ for a system of $N=2$ atoms with chirality $\Delta\gamma=0.8\gamma$ and laser detuning $\delta=\gamma/10$ as a function of the driving strength $\Omega$ and feedback pulse area $g$. Rates and fluctuations are given in units of $\gamma$.}\label{fig3}
\end{figure}

In the case of photon counting feedback each photon detection triggers a fast modulation of the excitation laser, realizing a pulse of area $g$. We start again by investigating a chain of $N=2$ atoms. In Fig. \ref{fig3}a and b, we show the photon count rate $k$ and fluctuations ${\Delta k}^2$ as a function of the bare, i.e. non-feedback, excitation laser Rabi frequency $\Omega$ and feedback pulse area $g$. One can clearly observe that the most apparent feature occurs at $g=\pi/2$, where a sharp enhancement of the emission rate and the fluctuations with respect to the non-feedback case takes place. Here, the feedback triggers a $\pi$-pulse for each atom immediately after the detection of a photon. Generally, this leads to an increase of the number of excited atoms and thus to an increase in the photon emission rate.

\begin{figure}
\includegraphics[width=\columnwidth]{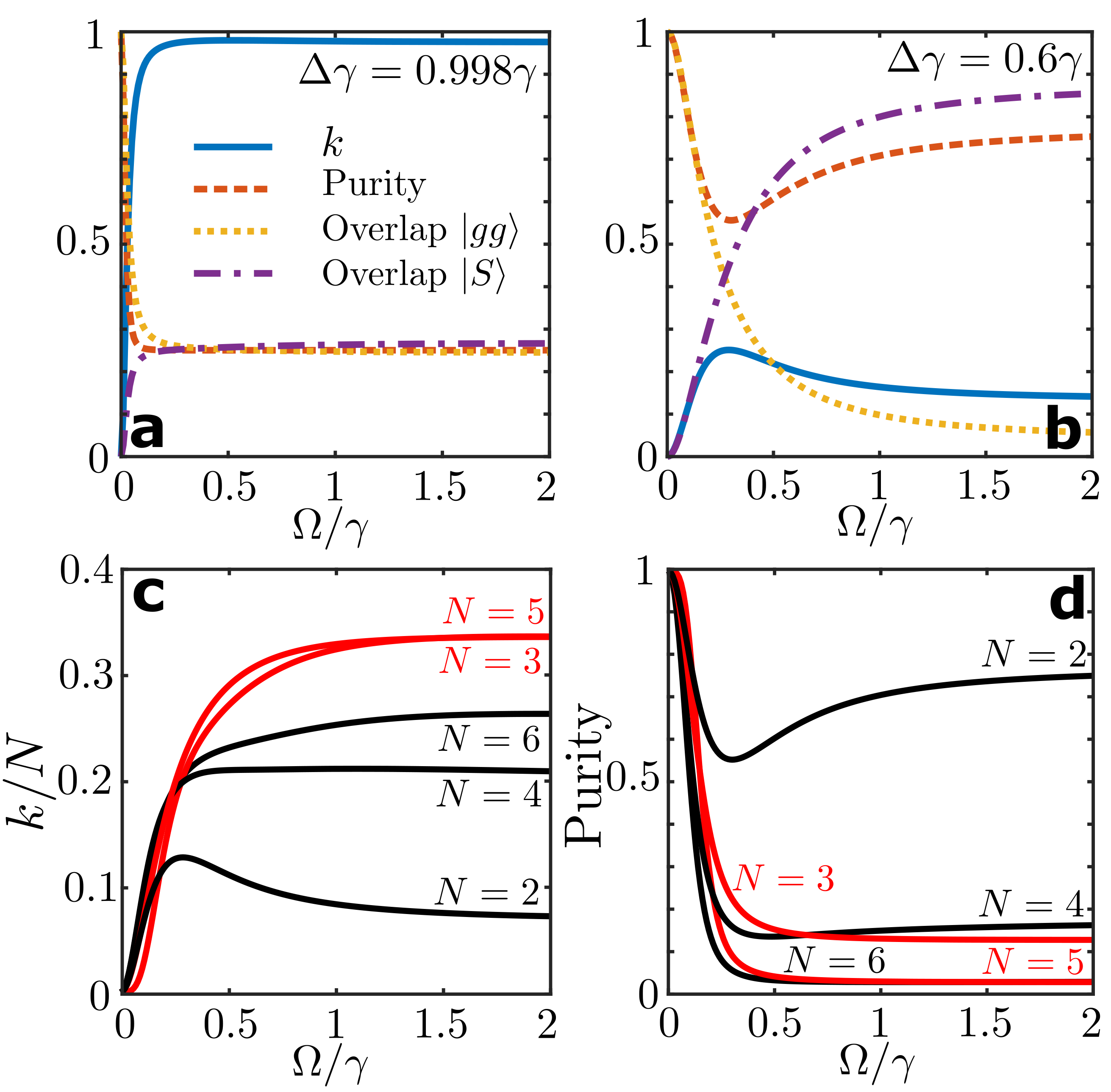}
\caption{\textbf{Chirality and finite size effects.} Photon counting rate $k$, purity and overlap of the stationary state with $\left|gg\right>$ and $\left|S\right>$ as a function of $\Omega$ for $N=2$ atoms. The area of the feedback pulse is $g=\pi/2$ and the detuning of the laser is $\delta=\gamma/10$ for \textbf{a:} almost perfect chirality, $\Delta\gamma=0.998\gamma$, and \textbf{b:} $\Delta\gamma=0.6\gamma$. \textbf{c:} Normalized photon counting rate and \textbf{d:} purity for a system formed by $N=2,3\dots 6$ atoms with $\Delta\gamma=0.6\gamma$. The black (red) lines represent even (odd) atom number $N$. Rates are given in units of $\gamma$.}\label{fig4}
\end{figure}

It is interesting to investigate the case $g=\pi/2$ at the level of the feedback master equation \eqref{countingfb}. Here, the jump operator $J_R=e^{-\mathrm{i}\frac{\pi}{2}(J+J^\dag)}J$ associated with the emission into the right-propagating mode becomes hermitian or antihermitian depending on the parity of the number of atoms, i.e. $J_R=(-1)^N J_R^\dag$. As a consequence, the stationary state becomes fully mixed when $\gamma_R=\Delta\gamma =\gamma$, i.e. when the emission of photons takes place solely into the right-propagating mode. This is illustrated in Fig. \ref{fig4}a, where we show the emission rate and the purity and overlap of the stationary state with the states $\left|gg\right>$ and $\left|S\right>$. The latter are consistent with a diagonal density matrix for $N=2$. For non-perfect chirality ($\Delta\gamma < \gamma$), the second decay channel (left-propagating mode) is open, and hence the fully mixed state ceases to be the stationary state. This is seen in Fig. \ref{fig4}b, which shows results similar to the ones depicted in Fig. \ref{fig2}b in the absence of feedback. The main difference is that with feedback the contributions of the states $\left|T\right>$ and $\left|ee\right>$ to the stationary state, and hence also the photon counting rate, are higher.

Also with feedback, the choice of the laser detuning $\delta$ has a strong impact on the stationary state of the system. As discussed previously, for even atom number $N$ and on resonance, $\delta=0$, the stationary state of the system is exactly the dimer state. This stationary state is unperturbed by the feedback, except at $g=\pi/2$ and $\Delta\gamma=\gamma$. At this particular point, both the fully mixed and the pure dimer states are stationary states of the system. The initial conditions of the system determine here the relative weight of the two states in the stationary state. However, as we move away from resonance, we find that at $g=\pi/2$ both states coexist \cite{buonaiuto2019}. The result is a dramatic increase of the fluctuations $\Delta k^2$, as shown in \ref{fig3}b. Finally, note that, since the dark dimer state only exists when the chain has an even atom number, for odd $N$ the stationary state at large values of $\Omega$ is simply the completely mixed state (see Figs. \ref{fig4}c and d).

\subsection{Homodyne feedback}
\begin{figure}
\includegraphics[width=\columnwidth]{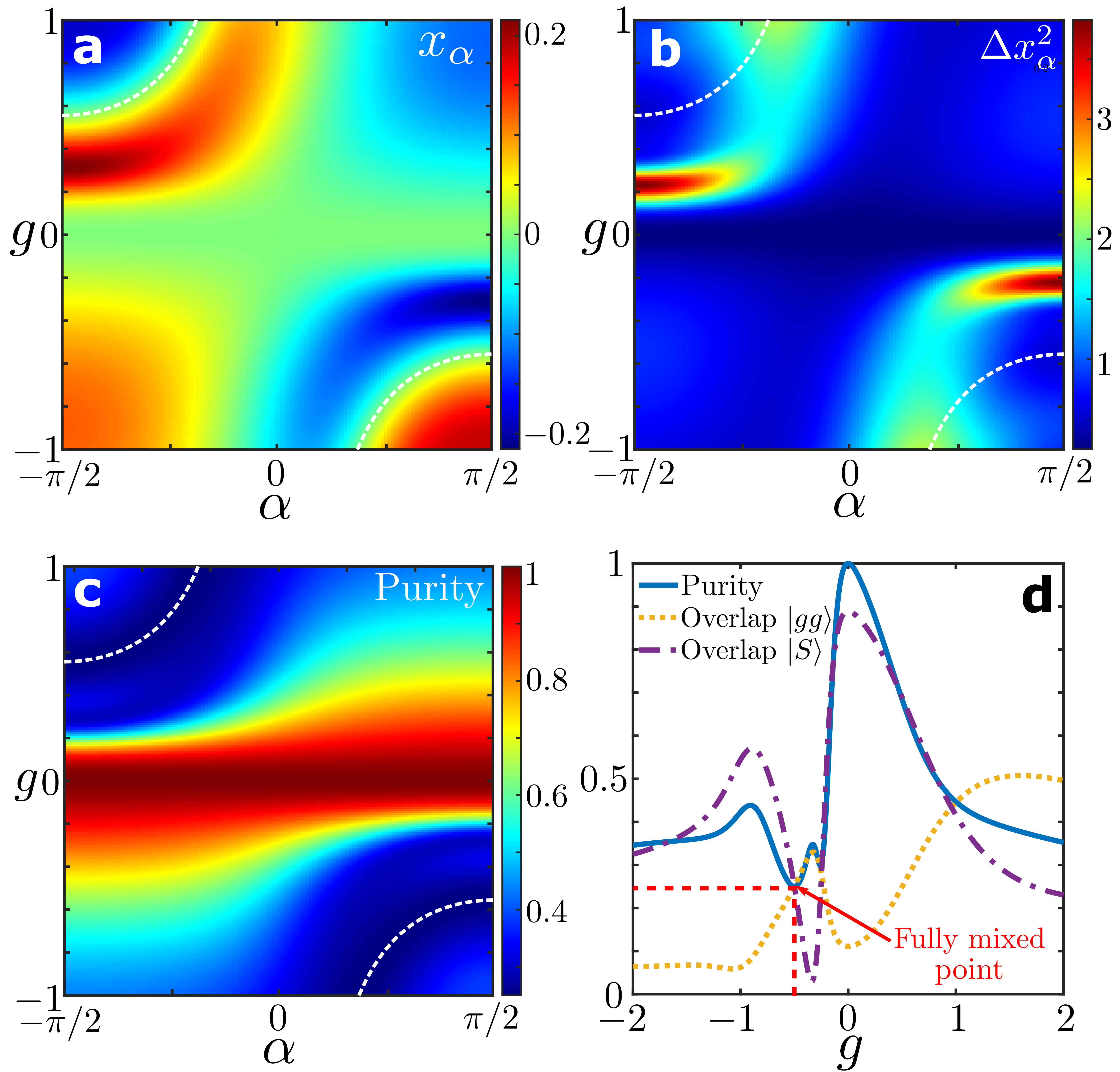}
\caption{\textbf{Homodyne feedback.} \textbf{a:} Average photocurrent quadrature per unit time $x_\alpha$, \textbf{b:} fluctuations $\Delta x_\alpha^2$, and \textbf{c:} purity of the stationary state for a system formed by $N=2$ atoms with parameters $\Omega=\gamma$, $\Delta\gamma=0.8\gamma$ and $\delta=0$. $x_\alpha$ and $\Delta x_\alpha^2$ are measured in units of $\gamma^{3/2}$ and $\gamma^{2}$, respectively. The white dashed lines represent the parameter region where the stationary state is the fully mixed state, given by Eq. \eqref{mix}. \textbf{d:} Purity and overlap of the stationary state with the states $\left|gg\right>$ and $\left|S\right>$ as a function of $g$ for $\Omega=\gamma$, $\Delta\gamma=\gamma$, $\delta=0$ and $\alpha=\pi/2$. The point where the fully mixed state is the stationary state, according to Eq. \eqref{mix}, is indicated.}\label{fig5}
\end{figure}

For homodyne feedback the dynamics is governed by the ME (\ref{homodynefb}). In Figs. \ref{fig5}a, b, and c we show the average and fluctuations of the photocurrent quadrature per unit time, $x_\alpha$ and $\Delta x_\alpha^2$, respectively, as well as the purity of the stationary state for $N=2$. One can see that along the $g=0$ line (in the absence of feedback) the purity of the stationary state is $1$, and both the average and fluctuations of $x_\alpha$ are zero. This is consistent with the pure dimer state $\left|D\right>$ being the stationary state. One can also observe that for $\alpha=0$, i.e. when the feedback is triggered by the photocurrent amplitude $X(t)$, the three observables change rather weakly, unlike for larger values of the quadrature angle $\alpha$, where the changes are sharper. In order to understand this, let us remember that the feedback alters the jump operator of photon emission into the right-propagating mode and the Hamiltonian as follows:
\begin{eqnarray*}
\sqrt{\gamma_R}J&\rightarrow& \sqrt{\gamma_R}\left[\left(1-\mathrm{i} g e^{\mathrm{i}\alpha}\right)J-\mathrm{i} g e^{\mathrm{i}\alpha}J^\dag\right],\\
H&\rightarrow& H+\frac{g\gamma_R}{2}\left(2\cos{\alpha}J^\dag J+e^{-\mathrm{i}\alpha}JJ+e^{\mathrm{i}\alpha}J^\dag J^\dag\right).
\end{eqnarray*}
Using these equations we can identify parameter combinations which, like in the photon counting feedback case, lead to an antihermitian right-propagating jump operator:
\begin{equation}\label{mix}
    g\left(\frac{\Delta\gamma}{\gamma}+1\right)=-\frac{1}{\sin{\alpha}}.
\end{equation}
Within this parameter space, the stationary state of the system is the fully mixed state for all values of the driving $\Omega$ and detuning $\delta$. This ''fully mixed line'', which is also independent of the number of atoms $N$, is depicted in Figs. \ref{fig5}a, b, and c. One can see here that in the vicinity of this line the properties of the emission and of the stationary state vary sharply as a function of $g$ for each fixed quadrature angle $\alpha$.

To keep the following analysis of the stationary state simple, we focus on $\Delta\gamma=\gamma$ and quadrature angle $\alpha=\pi/2$, with the corresponding data for the purity and overlap of the stationary state with $\left|gg\right>$ and $\left|S\right>$ shown in in Fig. \ref{fig5}d. Here, the feedback is triggered by the measurement of the photocurrent phase quadrature, $P(t)$, the Hamiltonian becomes
\begin{eqnarray*}
H_{\alpha=\pi/2}&=&\Omega\left(J+J^{\dagger}\right)-\frac{\mathrm{i}}{2}\gamma \sum_{j>l}(\sigma^{+}_{j}\sigma^{-}_{l}-\sigma^{+}_{l}\sigma^{-}_{j})\\
&&+\frac{\mathrm{i}}{2}g\gamma\left(J^\dag J^\dag-JJ\right).
\end{eqnarray*}
and the jump operator is modified as
\begin{equation}
\sqrt{\gamma_R}J\rightarrow \sqrt{\gamma_R}\left[\left(1+g\right)J+g J^\dag\right].
\end{equation}
When $|g|\gg1$ this operator is approximately hermitian, and accordingly the purity of the state decays eventually from 1 at $g=0$ to its minimum possible value, $1/2^N$. Unlike for $g>0$, the approach of the purity to this limit is non-monotonic for negative $g$, as can be seen in Fig. \ref{fig5}d. This is due to the existence of the ''fully mixed point'' given by Eq. \eqref{mix} located at $g=-1/2$, at which the modified jump operator is antihermitian. Similarly, the overlap of the stationary state with the many-body states $\left|gg\right>$ and $\left|S\right>$ undergo sharp changes in the vicinity of the fully mixed point.

\section{Conclusions and outlook} \label{sec:conclusions}

We have explored measurement-based feedback in a waveguide QED system, considering protocols based on photon counting and homodyne detection. The feedback consisted of a modulation of the driving laser field, which excites atoms that are chirally coupled to a nanofiber. In the future it will be interesting to explore whether a different choice of feedback operation, such as the action of a magnetic field, or a magnetic field gradient, can add further handles for controlling the photon emission or the properties of the atomic many-body state. 
In particular, it will be of interest to find protocols that allow to bring the atoms into an entangled state (beyond the dimer state), and to identify strategies that result in a photon output with desired properties, such as a regular train of photons or squeezed light. 

\begin{acknowledgements}
The research leading to these results has received funding from the European Union’s H2020 research and innovation  programme [Grant  Agreement No. 800942 (ErBeStA)] and EPSRC [Grant No. EP/R04340X/1]. IL acknowledges support from the ”Wissenschaftler-R\"uckkehrprogramm GSO/CZS of the Carl-Zeiss-Stiftung and the German Scholars Organization e.V.. BO was supported by the Royal Society and EPSRC [Grant No. DH130145]
\end{acknowledgements}

%

\end{document}